\documentclass{article}
\usepackage{graphics}
\begin{document}
\title{Maximizing bandgaps in two-dimensional photonic crystals: a variational algorithm}
\author{Prabasaj Paul\\Department of Physics and Astronomy, Colgate University,\\Hamilton NY 13346 \and Francis C. Ndi\\Physics Department, Lehigh University,\\ 16 Memorial Drive East, Bethlehem PA 18015}
\maketitle
\begin{abstract}
We present an algorithm for the maximization of photonic bandgaps in two-dimensional crystals. Once the translational symmetries of the underlying structure have been imposed, our algorithm finds a global maximal (and complete, if one exists) bandgap. Additionally, we prove two remarkable results related to maximal bandgaps: the so-called `maximum contrast' rule, and about the location in the Brillouin zone of band edges.
\end{abstract}
\section{Introduction}
Photonic crystals are structures where the electromagnetic properties of the constituent material are periodic in space. They have been intensely studied recently \cite{website} both for their practical applications and the theoretical challenges they  pose. One area of research is the design and fabrication of structures that do \emph{not} allow electromagnetic waves with frequencies in a certain range to propagate in them. Such structures are expected to find use as perfect mirrors and guiding structures, etc. In many applications, it is desirable that the excluded frequency range -- the bandgap -- be as large as possible. In this paper, we focus our attention on the design of structures with such maximal bandgaps.

The design problem that we propose to solve has been addressed before \cite{exact,numerical,plihal}. The starting point in each approach has been to impose the translational symmetries of the structure to be designed, and to impose global constraints that the material properties must staisfy (maximum and minimum dielectric constants, for instance). We do the same. In earlier approaches, the next step involved exploring a subset of the structures that satisfy the imposed translational symmetry. Our algorithm, in contrast, involves an unrestricted exploration of all the structures that satisfy the imposed conditions. Notably, the algorithm is polarization non-specific; it can maximize a complete bandgap (i.e. a frequency range that is excluded for all polarizations).

The algorithm we propose is based on incremental steps, each of which increases the size of the selected bandgap. Each step is based on a variational argument that is discussed in \cite{bs} and outlined below.

The plan of the paper is as follows. In the next section, we present the equations relevant to the  problem. Thereafter, we outline the variational argument we use and describe an algorithm to implement this in the case where the dielectric constant of the material is required to lie in a certain range. We have numerically implemented the algorithm; the details of the implementation will be presented elsewhere.

Two remarkable results follow quite easily from our discussion: First, that for a maximal bandgap structure where the dielectric constant is allowed to vary within a range, the dielectric constant at each point takes on either the smallest or the largest possible value. Second, that the wavevectors corresponding to the edges of a maximal bandgap may take on values only from a very small subset of the Brillouin zone.

\section{Maxwell equations in two-dimensional systems}
We consider a two-dimensional photonic crystal with a dielectric constant $\epsilon(\vec{\mathbf r})$ that is periodic in two independent directions (that define the $x$-$y$ plane) and homogeneous in the third (the $z$-direction). The magnetic susceptibility of the constituent material is assumed spatially constant. In this system, electromagnetic waves with wave vectors in the $x$-$y$ plane come in two distinct species: $E$-polarization, where the electric field vector is parallel to the $z$-direction; and, $H$-polarization, where the magnetic field vector is parallel to the $z$-direction. In the former case, Maxwell equations yield
\begin{equation}
\epsilon^{-1}\left(\frac{\partial^2}{\partial x^2}+\frac{\partial^2}{\partial y^2}\right)E+\frac{\omega^2}{c^2}E=0,
\end{equation} 
and in the latter,
\begin{equation}
\left(\frac{\partial}{\partial x}\epsilon^{-1}\frac{\partial}{\partial x}+\frac{\partial}{\partial y}\epsilon^{-1}\frac{\partial}{\partial y}\right)H+\frac{\omega^2}{c^2}H=0.
\end{equation}
$E$ and $H$ are the $z$-components of the electric field and magnetization vectors, respectively. (See \cite{plihal} for details of derivation.)
Each of these equations has the form $Du=\lambda u$, where $D$ is a Hermitian operator. In the case of $E$-polarization, 
\begin{equation}
D\equiv\epsilon^{-1/2}\left(\frac{\partial^2}{\partial x^2}+\frac{\partial^2}{\partial y^2}\right)\epsilon^{-1/2}
\end{equation} 
and $u\equiv\epsilon^{1/2}E$. For $H$-polarization, \begin{equation}
D\equiv\left(\frac{\partial}{\partial x}\epsilon^{-1}\frac{\partial}{\partial x}+\frac{\partial}{\partial y}\epsilon^{-1}\frac{\partial}{\partial y}\right)
\end{equation}
and $u\equiv H$. In both cases, $\lambda\equiv -\omega^2/c^2$. For ease of exposition, we will set $c=1$.

It is well-known that the spectra of eigenvalues of the equations above display `bands'. Given two bands, the difference between the smallest eigenvalue of the upper band and the largest eigenvalue of the lower band characterizes the bandgap. (We will focus our attention on the spectrum of $\omega$, rather than $\lambda$.) After having selected two adjacent bands, we will seek a function $\epsilon(\vec{\mathbf r})$ that maximizes the bandgap between them. In general, $\epsilon(\vec{\mathbf r})$ may be subject to imposed constraints. While the method outlined below may be used to handle a variety of constraints, we will confine ourselves to a specific one: $\epsilon_{min}\le\epsilon(\vec{\mathbf r})\le\epsilon_{max}$.
\section{The variational algorithm}
Starting with the standard eigenvalue problem $Du=\lambda u$, where $D$ is a Hermitian operator, we will first investigate changes in $\lambda$ due to changes in $D$. Prefixing $\delta$ to denote `change in'
$$(u+\delta u)^\dagger(D+\delta D)(u+\delta u)=\lambda+\delta\lambda.$$
Retaining terms to first order in the changes, and noting that for normalized eigenfunctions, $u^\dagger\delta u=0$,
$$u^\dagger\delta Du=\delta\lambda.$$ 
Change in $D$ is due to change $\delta\epsilon$ in $\epsilon$. In the case of $E$-polarization,
\begin{eqnarray}
\delta D&=&\delta\left[\epsilon^{-1/2}\left(\frac{\partial^2}{\partial x^2}+\frac{\partial^2}{\partial y^2}\right)\epsilon^{-1/2}\right]\nonumber\\
&=&-\frac{\delta\epsilon}{2\epsilon}D-D\frac{\delta\epsilon}{2\epsilon}
\end{eqnarray}
so that 
\begin{equation}
u^\dagger\delta Du=-\lambda\int_Cu^*\frac{\delta\epsilon}{\epsilon}u\,d\tau,
\end{equation}
where we use $d\tau$ to denote the volume element of the unit cell $C$ with boundary $S$.

One may proceed similarly in the case of $H$-polarization. Using the notational shorthand $\vec\nabla\equiv\hat{\mathbf i}\,\partial/\partial x+\hat{\mathbf j}\,\partial/\partial y$, 
\begin{eqnarray}
u^\dagger\delta Du&=&\int_C u^*\vec\nabla\cdot(\delta(\epsilon^{-1})\vec\nabla u)\,d\tau\nonumber\\
&=&\int_S(u^*\delta(\epsilon^{-1})\vec\nabla u)\cdot\hat{n}\,dA-\int_C(\vec\nabla u^*)\cdot(\vec\nabla u)\delta(\epsilon^{-1})\,d\tau\nonumber\\
&=&-\int_C\delta(\epsilon^{-1})|\vec\nabla u|^2\,d\tau.
\end{eqnarray} 

Recall that $\delta\lambda=-2\omega\delta\omega$. To sum up, we have the following:
\begin{equation}
\label{delOdelE}
\frac{\delta\omega}{\delta\epsilon}=\left\{\begin{array}{r@{\quad:\quad}l} {\omega|u|^2}/{2\epsilon} & E-\mbox{polarization}\\
-{|\vec{\nabla}u|^2}/{2\omega\epsilon} & H-\mbox{polarization.}\end{array}\right.
\end{equation}

It is not $\omega$ but, rather, a difference $\omega_2-\omega_1$, that we seek to maximize with respect to variations in $\epsilon$. Thus, we would like to tailor variations in $\epsilon$ so that $\delta(\omega_2-\omega_1)>0$. This is achieved if $\delta\epsilon>0$ wherever $\delta\omega_2/\delta\epsilon>\delta\omega_1/\delta\epsilon$, and vice versa. If $\epsilon=\epsilon_{max}$ ($\epsilon=\epsilon_{min}$), it is impossible to achieve $\delta\epsilon>0$ ($\delta\epsilon<0$). Therefore, maximal $\omega_2-\omega_1$ is expected when $\epsilon=\epsilon_{max}$ wherever $\delta\omega_2/\delta\epsilon>\delta\omega_1/\delta\epsilon$, and $\epsilon=\epsilon_{min}$ wherever $\delta\omega_2/\delta\epsilon<\delta\omega_1/\delta\epsilon$. Note that no explicit reference to polarization has been made; $\omega_1$ and $\omega_2$ may even be eigenvalues corresponding to different polarizations. This makes it possible for our algorithm to maximize the size of complete bandgaps.

This suggests the following iterative algorithm to a maximal bandgap:
\begin{itemize}
\item[\bf 0.] Choose periodicity of $\epsilon(\vec{\mathbf r})$, and the adjacent bands that straddle the bandgap to be maximized. Initialize with an arbitrary (but appropriately periodic) $\epsilon(\vec{\mathbf r})$.
\item[\bf 1.] Determine, using $\epsilon(\vec{\mathbf r})$, the spectrum of eigenvalues $\omega$ and the wavefunctions $u$.
\item[\bf 2.] Determine the wavefunctions that correspond to the smallest eigenvalue of the upper band ($u_2$ and $\omega_2$) and the largest eigenvalue of the lower band ($u_1$ and $\omega_1$).
\item[\bf 3.] Make \emph{small} changes $\delta\epsilon$ in $\epsilon(\vec{\mathbf r})$: $\delta\epsilon\ge0$ wherever $\delta\omega_2/\delta\epsilon>\delta\omega_1/\delta\epsilon$ (determined using the expressions in (\ref{delOdelE})), and vice versa. (Maintain consistency with the constraint $\epsilon_{min}\le\epsilon(\vec{\mathbf r})\le\epsilon_{max}$.)
\item[\bf 4.] Go to {\bf 1} unless termination/convergence criteria are met.  
\end{itemize}
A discussion of some of the finer points of each step follows.

The initialization step requires the lattice parameters as input, both to establish a length scale and to fix the translational symmetry of the system. It is in this sense -- and this sense only -- that the maximum attained is not global. (Observe that, due to the scaling properties of the eigenvalue equations, maximization without an imposed length scale is trivial.)

Considerable work \cite{num,fp} has been done towards numerical solution of the eigenvalue equations for arbitrary $\epsilon(\vec{\mathbf r})$. The usual practice is to first discretize the space in some manner, then select a reasonably dense subset of wavevectors from the Brillouin zone and, finally, obtain the eigenvalues corresponding to each wavevector. 

Remarkably, it turns out that it is not necessary to perform a complete bandstructure computation to implement our scheme. To see this, note first that, since $\epsilon(\vec{\mathbf r})$ is real, wavevectors $\pm\vec{\mathbf k}$ have degenerate spectra. Therefore, any arbitrary linear combination $u=\alpha u_{\vec{\mathbf k}}+\beta u_{-\vec{\mathbf k}}$ is an eigenfunction. Note, too, from our earlier discussion (equation (\ref{delOdelE}) and following paragraph), that either of the  wavefunctions $u=u_2$ or $u=u_1$ that straddle the bandgap must be such that $|u|^2$ (or $|\vec\nabla u|^2$) has the same translational symmetry as the underlying lattice. This severely limits the possible values of $\vec{\mathbf k}$ at the band-edges. In the case of a square lattice, it is easy to see that the only possible values of $\vec{\mathbf k}$ are those with $0$ or $\pm\pi$ as components. It is, thus, possible to limit the search -- for a maximal-bandgap producing configuration of $\epsilon(\vec{\mathbf r})$ in a 2D square lattice -- to wavevectors $0$, $(0,\pi)$, $(\pi,0)$ and $(\pi,\pi)$!

Another remarkable result -- referred to following equation (\ref{delOdelE}) and hinted at in step {\bf 3} of the algorithm -- is the`maximum-contrast' rule: if $\epsilon(\vec{\mathbf r})$ is allowed to vary arbitrarily between two limits then maximal bandgaps are obtained when $\epsilon(\vec{\mathbf r})$ attains one or the other extremal value at each point. (See \cite{bs} for a related detailed discussion.) 

One question that we have not answered here is whether the iterative algorithm converges. In a one-dimensional system (where exact results may be obtained otherwise and where $E$- and $H$-polarization spectra are degenerate), the algorithm does indeed converge rapidly to expected results. In 2D, numerical results obtained so far provide very strong evidence in favor of convergence. These results will be presented elsewhere \cite{fp}.

\end{document}